\documentclass[aps,twocolumn,superscriptaddress,prl,floatfix]{revtex4}

\usepackage{graphicx}
\usepackage{amsmath}
\usepackage{amssymb}
\usepackage{units}
\usepackage{times}

\renewcommand{\vec}[1]{\boldsymbol{#1}}

\begin{document}

\title{Tuning the spin texture in binary and ternary surface alloys on Ag(111)}

\author{Isabella Gierz}
\email[Corresponding author; electronic address:\
]{i.gierz@fkf.mpg.de}
\affiliation{Max-Planck-Institut f\"ur Festk\"orperforschung, D-70569 Stuttgart, Germany}
\author{Fabian Meier}
\affiliation{Physik-Institut, Universit\"at Z\"urich, 8057
Z\"urich, Switzerland}
\affiliation{Swiss Light Source, Paul
Scherrer Institute, 5232 Villigen, Switzerland}
\author{J.~Hugo Dil}
\affiliation{Physik-Institut, Universit\"at Z\"urich, 8057
Z\"urich, Switzerland}
\affiliation{Swiss Light Source, Paul
Scherrer Institute, 5232 Villigen, Switzerland}
\author{Klaus Kern}
\affiliation{Max-Planck-Institut f\"ur Festk\"orperforschung, D-70569 Stuttgart, Germany}
\affiliation{Institut de Physique de la Mati{\`e}re Condens{\'e}e, Ecole Polytechnique F{\'e}d{\'e}rale de Lausanne, CH-1015 Lausanne, Switzerland}
\author{Christian R. Ast}
\affiliation{Max-Planck-Institut f\"ur Festk\"orperforschung, D-70569 Stuttgart, Germany}

\date{\today}

\begin{abstract}
Recently, a giant spin splitting has been observed in surface alloys on noble metal (111) surfaces as a result of a strong structural modification at the surface as well as the large atomic spin-orbit interaction (SOI) of the alloy atoms. These surface alloys are an ideal playground to manipulate both the size of the spin splitting as well as the position of the Fermi level as it is possible to change the atomic SOI as well as the relaxation by varying alloy atoms and substrates. Using spin- and angle-resolved photoemission spectroscopy in combination with quantitative low energy electron diffraction we have studied the mixed binary Bi$_x$Sb$_{1-x}$/Ag(111) and the mixed ternary Bi$_x$Pb$_y$Sb$_{1-x-y}$/Ag(111) surface alloys where we observed a continuous evolution of the band structure with $x$ and $y$.
\end{abstract}

\maketitle

\section{Introduction}

Spin-degeneracy is the consequence of both time reversal and spatial inversion symmetry. The latter is broken at interfaces (surfaces) so that the two-dimensional (2D) states localized at this interface (surface) become spin-polarized in the presence of spin-orbit interaction (SOI). The Rashba-Bychkov (RB) model \cite{Rashba} describes how the spin-degeneracy in an asymmetrically confined 2D electron gas is lifted. This Rashba spin splitting depends on the magnitude of a potential gradient $\nabla V$ perpendicular to the confinement plane of the 2D electron gas. The possibility to tune the spin splitting by an external gate voltage forms the basis for many spintronic device proposals such as the Datta-Das spin field effect transistor \cite{Datta}. Furthermore, a gradient in the effective magnetic field caused by a spatial variation of the Rashba-type spin splitting leads to spin separation in the Stern-Gerlach spin filter \cite{Ohe}. In addition, these 2D spin-polarized states show the intrinsic spin Hall effect \cite{Sinova,Wunderlich} and an enhancement of the superconducting transition temperature is predicted in the regime where the Rashba splitting is larger than the Fermi energy \cite{Cappelluti}.

The RB model also gives a qualitative description for spin-polarized states localized at the surfaces of heavy metals such as Au(111) \cite{LaShell,Reinert01,Henk}, W(110) \cite{Rotenberg}, Bi(111), Bi(110), Bi(100) \cite{Koroteev} and Sb(111) \cite{Sugawara}. There, however, the potential gradient alone cannot explain the size of the measured spin splitting. Decorating the surfaces with different atoms/adlayers modifies the Rashba-type spin splitting in the states at the surface, e.\ g.\ (Ar, Kr, Xe) on Au(111) \cite{Forster,Moreschini1}, Li on (W(110), Mo(110)) \cite{Rotenberg}, (Au, Ag) layers on (W(110), Mo(110)) \cite{Shikin}. The biggest increase of the spin splitting up to now was achieved by surface alloying, where every third atom in a noble metal (111) surface (Ag(111) \cite{Pacile,Ast1,Ast2,Moreschini2,Meier1,Meier2,Bentmann} or Cu(111) \cite{Moreschini3,Bentmann}) is replaced by a heavy alloy atom (Bi, Pb and Sb). These surface alloys exhibit a spin splitting that is orders of magnitude larger than, e.\ g.\ in semiconductor heterostructures \cite{Ast1}.

For future device applications, it is necessary to tune the size of the spin splitting as well as the position of the Fermi level in such Rashba systems so that the Fermi surface possesses the desired spin texture. A first step in this direction shows that the spin splitting (and at the same time the Fermi energy) was controlled via the Bi content in a mixed Bi$_x$Pb$_{1-x}$/Ag(111) surface alloy \cite{Ast3,Meier1}.

Here we show by (spin- and) angle-resolved photoemisson spectroscopy [(S)ARPES] and quantitative low energy electron diffraction [$I(V)$-LEED], that it is possible to form a well-ordered mixed binary Bi$_x$Sb$_{1-x}$/Ag(111) surface alloy where the spin splitting can be tuned while leaving the position of the Fermi level largely unaffected. This is an important step towards the realization of a mixed ternary Bi$_x$Pb$_y$Sb$_{1-x-y}$/Ag(111) surface alloy, where spin splitting and Fermi energy can be tuned independently by varying the material parameters $x$ and $y$. As a proof of principle we have grown the ternary Bi$_{0.3}$Pb$_{0.35}$Sb$_{0.35}$/Ag(111) surface alloy and investigated its band structure by (S)ARPES.

\subsection{Surface alloys on Ag(111)}

\begin{figure}
  \includegraphics[width = 1\columnwidth]{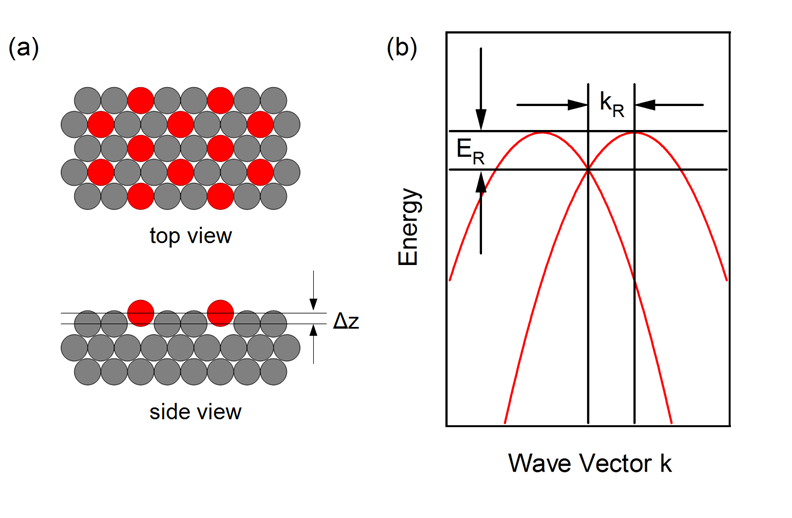}
  \caption{(Color online) (a) surface alloys on noble metal (111) surfaces form a $(\sqrt{3}\times\sqrt{3})R30^{\circ}$ reconstruction where every third substrate atom (grey) in the topmost layer is replaced by an alloy atom (red, dark grey). The alloy atoms relax outward by an amount $\Delta z$. (b) characteristic dispersion of the 2D quasi free electron gas with RB-type spin-orbit interaction. }
  \label{fig:figure1}
\end{figure}

\begin{table*}
\caption{Characteristic parameters for the $sp_z$ surface state of the different surface alloys on Ag(111).}
\label{Tab:table1}
\begin{tabular*}{\textwidth}{c@{\extracolsep\fill}ccccccc}
\hline\hline
&$\mathbf{\alpha_R}$ & $\mathbf{k_R}$ & $\mathbf{E_0}$ & $\mathbf{m^*}$ &  $\mathbf{\Delta z}$ & \textbf{Ref.}\\
\hline
\textbf{Bi/Ag(111)} & 3.2\,eV\AA & 0.13\AA$^{-1}$ & $-$0.135\,eV & $-$0.31\,m$_{\text{e}}$ & 0.65\AA & \cite{Ast1,Ast2,Gierz}\\
\textbf{Pb/Ag(111)} & 1.52\,eV\AA & 0.03\AA$^{-1}$ & +0.654\,eV & $-$0.15\,m$_{\text{e}}$ & 0.46\AA & \cite{Pacile,Ast2,Gierz}\\
\textbf{Sb/Ag(111)} & 0.38\,eV\AA & 0.005\AA$^{-1}$ & $-$0.28\,eV & $-$0.10\,m$_{\text{e}}$ & 0.10\AA & \cite{Moreschini2,Meier2,Gierz} \\
\hline\hline
\end{tabular*}
\end{table*}

The three surface alloys Bi/Ag(111), Pb/Ag(111) and Sb/Ag(111) all form the same $(\sqrt{3}\times\sqrt{3})R30^{\circ}$ reconstruction with respect to the Ag(111) substrate. In contrast to the clean (111) substrate the surface of the surface alloy is corrugated due to the outward relaxation $\Delta z$ of the alloy atoms (see Fig. \ref{fig:figure1}a) \cite{Gierz}. Furthermore, all three surface alloys show a very similar surface state band structure consisting of two sets of spin-split bands: one at lower binding energy with mainly $sp_z$ orbital character, one at higher binding energy with mainly $p_{xy}$ orbital character. The typical RB-type dispersion for each set of spin-split states is given by $E(k_{||})=\frac{\hbar^2}{2m^*}(k_{||}\pm k_R)^2+E_0$, where $m^*$ is the effective mass and $E_0$ is the position of the band maximum. This is illustrated in Fig. \ref{fig:figure1}b where the momentum offset $k_R$ and the Rashba energy $E_R$ of the RB-type dispersion are indicated by arrows. The Rashba parameter $\alpha_R=\hbar^2 k_R/m^*$ is a measure for the size of the spin splitting and closely related to the Rashba energy $E_R=\hbar^2 k_R^2/2m^*$. The two parabolas are completely spin-polarized, with the spin-orientation perpendicular to both $k_{||}$ and $\nabla V$. As $\nabla V$ lies along the surface normal in the framework of the RB model the spin-polarization is completely in-plane and parallel to the circular constant energy contours.

The RB model was successfully applied to qualitatively describe the spin-split dispersion on different noble metal surfaces, but it fails to make accurate quantitative predictions concerning the size of the spin splitting. A more quantitative description can be obtained with a tight-binding calculation which explicitly includes the contribution of the atomic SOI \cite{Petersen}. The characteristic parameters for the lower $sp_z$ band for the three surface alloys (Bi,Pb,Sb)/Ag(111) are summarized in Table \ref{Tab:table1}. The spin splitting $k_R$ increases from Sb via Pb to Bi with increasing mass and increasing outward relaxation $\Delta z$ of the alloy atom. As Bi and Sb have the same number of valence electrons, i.\ e.\ they are isoelectronic, the band maximum $E_0$ is located at a similar energetic position in the occupied states for the Bi/Ag(111) and the Sb/Ag(111) surface alloys. Pb, however, has one valence electron less than both Bi and Sb. Therefore, the band maximum of the $sp_z$ state is in the unoccupied states. All the $sp_z$ states have negative effective masses ranging from -0.10m$_{\text{e}}$ for Sb/Ag(111) to -0.35m$_{\text{e}}$ for Bi/Ag(111), where m$_{\text{e}}$ is the free electron mass.

As only a fraction of the atoms in the surface alloys exhibit a significant atomic SOI, the observed giant spin splitting in surface alloys has to be sought in the structure itself. More specifically, a considerable outward relaxation of the alloy atoms determines the orbital composition of the surface state, which was shown to enhance the spin splitting \cite{Bihlmayer2,Gierz}. A corrugation of the surface structure will lead to an out-of-plane spin component \cite{Henk2}. In the free electron RB model this can be understood as originating from an in-plane component of $\nabla V$ resulting from an in-plane inversion asymmetry \cite{Premper}.

Mixing Bi and Pb in a binary Bi$_x$Pb$_{1-x}$/Ag(111) surface alloy leads to a continuous evolution of the characteristic parameters of the $sp_z$ surface state dispersion with Bi content $x$ \cite{Ast3,Meier1}. The spin splitting $k_R$ increases with increasing Bi content accompanied by a downward shift of the band maximum into the occupied states. Bi and Sb on the other hand are isoelectronic, i.e. mixing Bi and Sb in a binary Bi$_x$Sb$_{1-x}$/Ag(111) surface alloy would offer the possibility to tune the size of the spin splitting without changing the Fermi level. However, the Sb/Ag(111) surface alloy forms with either face centered cubic (fcc) or hexagonally closed packed (hcp) toplayer stacking depending on the growth conditions \cite{Gierz,Woodruff,Quinn}, whereas Bi/Ag(111) as well as Pb/Ag(111) always form with fcc toplayer stacking. Therefore, it is not {\it a priori} clear whether a well ordered Bi$_x$Sb$_{1-x}$/Ag(111) surface alloy can be formed.

\section{Experiments}

\subsection{Mixed binary Bi$_x$Sb$_{1-x}$/Ag(111) surface alloy}

\begin{figure*}
  \includegraphics[width = 2\columnwidth]{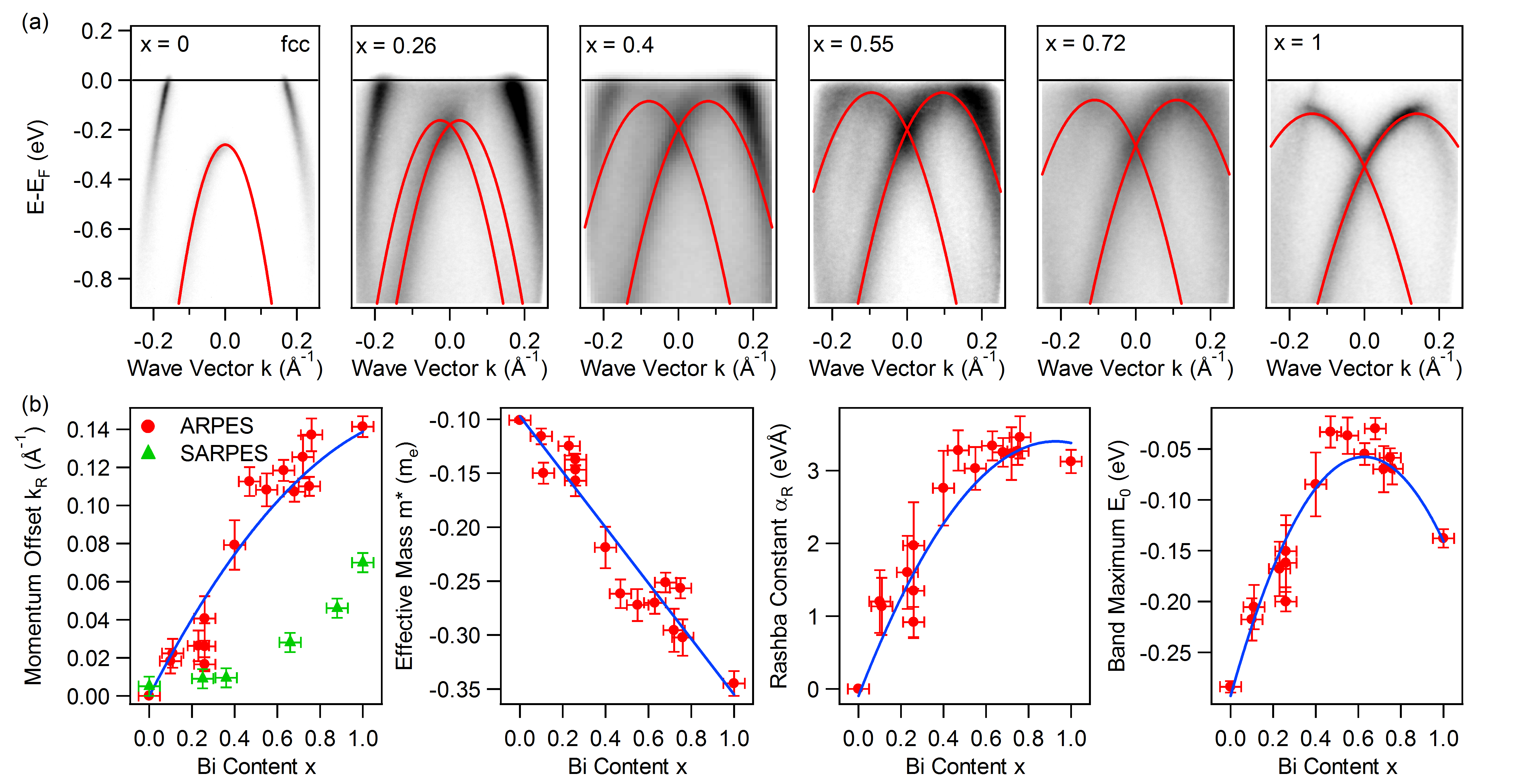}
  \caption{(Color online) Experimental photoemission from Bi$_{x}$Sb$_{1-x}$/Ag(111). The evolution of the $sp_{z}$ surface state dispersion is shown as a function of Bi content $x$ on a linear grey scale with black (white) corresponding to high (low) photocurrents (a). From parabolic fits to the data (red lines in (a)), the characteristic parameters of the Rashba model (Momentum offset $k_R$, effective mass $m^{\star}$, Rashba constant $\alpha_R$, and band maximum $E_0$) have been determined (b). Red dots and green triangles were obtained by spin-integrated and spin-resolved ARPES, respectively. Blue lines are guides to the eye.}
  \label{fig:ARPES}
\end{figure*}

All experiments were performed in ultra-high vacuum with a base pressure of $1 \times 10^{-10}$\,mbar. The ARPES experiments were done with a SPECS Phoibos 150 hemispherical analyzer with an energy resolution of 10\,meV and monochromatized He I radiation at $h\nu=21.2$\,eV. For the LEED measurements an ErLEED 1000-A was used. ARPES and LEED experiments were performed at liquid nitrogen temperature.

The spin splitting of the $sp_{z}$ surface state in Sb/Ag(111) is smaller than the line width of the bands which prevents its investigation by conventional spin-integrated ARPES \cite{Moreschini2}. The discrimination of the two bands, however, can be achieved by spin-resolved ARPES (SARPES), where the spin polarization $P$ of the photoelectrons is measured in addition to their kinetic energy and the emission angle \cite{Dil}. Due to the low efficiency of present Mott detectors SARPES measurements are very time-consuming and are therefore usually restricted to single spin-resolved momentum distribution curves (MDCs). The intensities for spin-up (spin-down) electrons $I^{\uparrow}$ ($I^{\downarrow}$) are obtained from the measured spin-integrated intensity $I_{tot}$ according to $I^{\uparrow} = (1 + P) I_{\mathrm{tot}} / 2$ and $I^{\downarrow} = (1 - P) I_{\mathrm{tot}} / 2$. Assuming a parabolic dispersion, the Rashba splitting $k_R$ is then given by $k_R=\Delta k/2$, where $\Delta k$ is the $\vec{k}_{\parallel}$-distance of the maxima in $I^{\uparrow}$ and $I^{\downarrow}$. Note that $k_R\neq\Delta k/2$ if the dispersion of the bands is not completely parabolic.

SARPES experiments were performed at the Surface and Interface spectroscopy beamline at the Swiss Light Source of the Paul Scherrer Institute using the COPHEE spectrometer \cite{hoesch:2002}. This spectrometer is equipped with two orthogonal Mott polarimeters, which can measure the spin expectation value for an arbitrary state in reciprocal space. The energy and angle resolution are 80\,meV and 1.5$^{\circ}$, respectively. The data were obtained using synchrotron radiation of 24\,eV at room temperature.

The Ag(111) substrate was cleaned using several sputtering-annealing cycles (sputtering with 1\,keV Ar ions at an Ar pressure of $1\times 10^{-6}$\,mbar followed by annealing at 530$^{\circ}$C). Cleanliness was controlled with X-ray photoemission spectroscopy (XPS). In addition the surface state of clean Ag(111) was monitored with ARPES. For the preparation of the mixed Bi$_{x}$Sb$_{1-x}$/Ag(111) surface alloy for the ARPES experiments, Sb and Bi were successively deposited on the Ag(111) substrate using a commercial electron beam evaporator. The substrate temperature was 250$^{\circ}$C during Sb deposition. For the subsequent Bi deposition the substrate temperature was reduced to 150$^{\circ}$C. For the SARPES experiments Bi and Sb were deposited simultaneously. After deposition the LEED pattern showed a sharp ($\sqrt{3}\times\sqrt{3}$)R30$^{\circ}$ structure.

In order to prepare the ternary surface alloy Bi$_{x}$Pb$_{y}$Sb$_{1-x-y}$/Ag(111), Sb was deposited first with the sample at 200$^{\circ}$C and then Bi and Pb were simultaneously deposited at 150$^{\circ}$C. The LEED pattern of the ternary surface alloy showed a spot broadening, which we attribute to a larger disorder.

The Bi, Sb, and Pb coverage was determined by analyzing the integrated intensity of the respective core levels, measured with XPS.

\section{Results and discussion}

Figure \ref{fig:ARPES}a shows the evolution of the surface state band structure for the mixed binary alloys Bi$_{x}$Sb$_{1-x}$/Ag(111) as a function of Bi content $x$ measured with ARPES. The dispersion of the $sp_{z}$ surface state evolves continuously between $x = 0$ and $x = 1$. Around $x = 0.5$ the line width of the bands increases considerably which we attribute mainly to structural disorder, i.e. to imperfections in the $(\sqrt{3}\times\sqrt{3})R30^{\circ}$ reconstruction. The position of the lower $sp_{z}$ band was determined by fitting MDCs with a Lorentzian and a constant background. The resulting $E(k)$ data were then fitted by parabolas to determine the characteristic Rashba parameters. The continuous evolution of momentum offset $k_R$, Rashba constant $\alpha_R$, band maximum $E_0$, and effective mass $m^{\star}$ with increasing Bi content $x$ is shown in Fig. \ref{fig:ARPES}b. The Rashba parameter $\alpha_R$ has been calculated from the experimentally determined momentum offset and effective mass.

While $k_R$, $m^{\star}$, and $\alpha_R$ continuously increase with $x$, the band maximum $E_0$ reaches a maximum at about $x \approx 0.63$ and then decreases again. It is known that $E_0$ is correlated with the outward relaxation of the Bi and Sb atoms \cite{Bihlmayer2}. Because it is unlikely that the outward relaxation of the alloy atoms in the mixed Bi$_x$Sb$_{1-x}$/Ag(111) surface alloy is larger than in the pure Bi/Ag(111) surface alloy we attribute the maximum at $x\approx 0.63$ to structural disorder, which is corroborated by the considerable increase in line width.

Recent first principles calculations are in good agreement with our experimental values for $k_R$ \cite{Mirhosseini}. However, in contrast to experiment the calculations show a continuous increase of $E_0$ with Bi concentration $x$ and do not capture the maximum at $x \approx 0.63$. This is to be expected because structural disorder is not included in the calculations.

\begin{figure}
\includegraphics[width = 1\columnwidth]{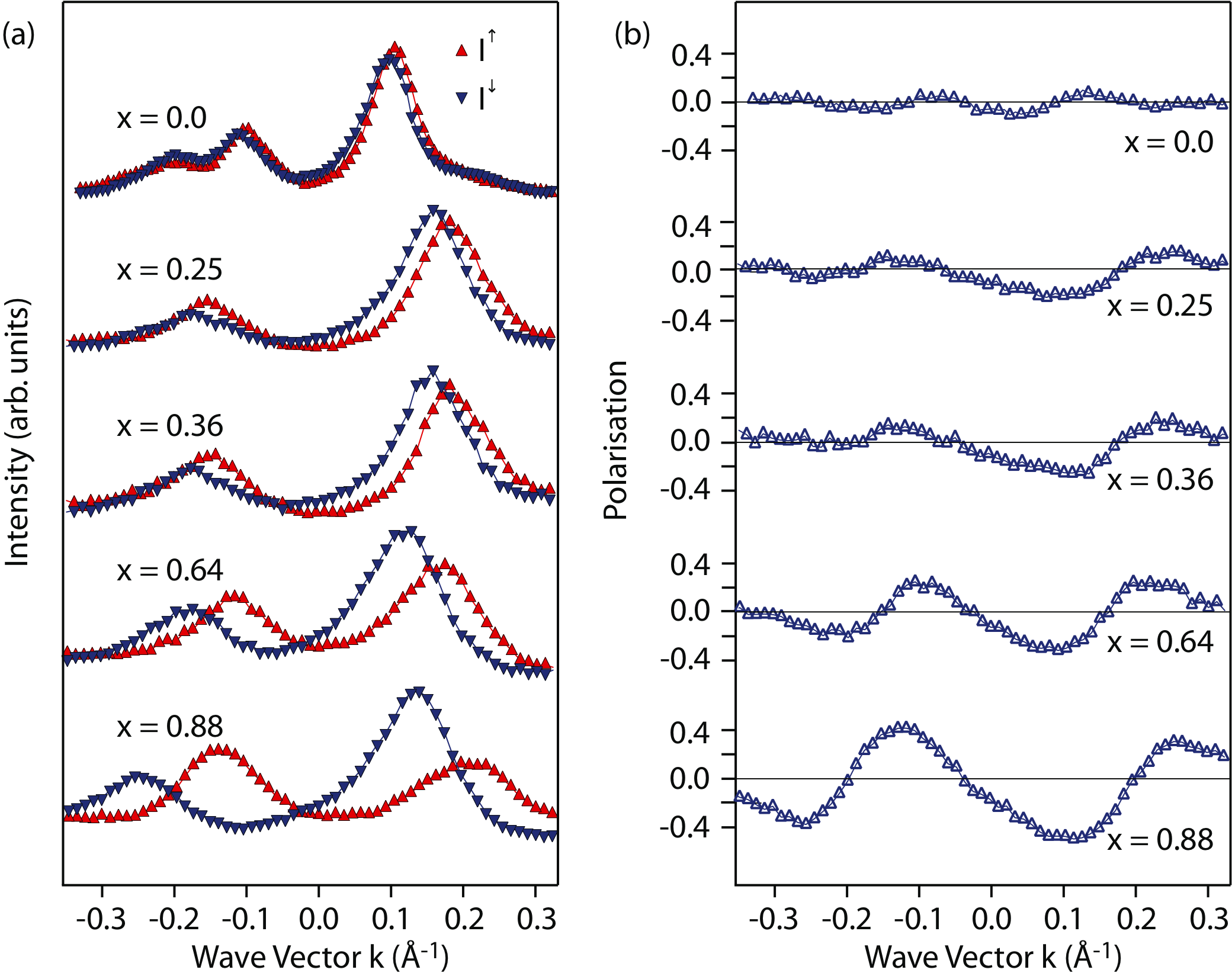}
\caption{(Color online) Spin-resolved photoemission from the mixed Bi$_{x}$Sb$_{1-x}$/Ag(111) surface alloy. Intensities $I^{\uparrow}$ (red triangles pointing upwards) and $I^{\downarrow}$ (blue triangles pointing downwards) (a) and spin-polarization (b) of momentum distribution curves from Bi$_{x}$Sb$_{1-x}$/Ag(111) at an initial state energy of -0.6\,eV as a function of Bi content $x$.}
\label{fig:SARPES1}
\end{figure}

For small Bi contents $x$ the spin splitting of the $sp_{z}$ surface state of Bi$_x$Sb$_{1-x}$/Ag(111) is comparable to the line width of the bands. This limits the accuracy of the values for the spin splitting obtained by conventional spin-integrated ARPES. To overcome this problem Fig. \ref{fig:SARPES1} shows spin-resolved MDCs recorded at an initial state energy of $-$0.6\,eV as a function of Bi content $x$ for the mixed Bi$_{x}$Sb$_{1-x}$/Ag(111) surface alloy. The SARPES-derived momentum offset $\Delta k /2$ is included in Fig. \ref{fig:ARPES}b (green triangles).

For a Bi content $x \gtrsim 0.3$ the $sp_z$-band is not parabolic any longer because it hybridizes with the upper $p_{xy}$-band. In this case, the values for $k_R$ as determined by ARPES and $\Delta k/2$ as determined by SARPES deviate systematically. Nevertheless, the discrepancy for low values of $x$, where the dispersion is expected to be parabolic, is unexpectedly large. Although we cannot fully explain this discrepancy we want to point out that for small $x$ a determination of $k_R$ with conventional spin-integrated ARPES is difficult for several reasons. First, the size of the spin splitting is comparable to the line width of the bands, so that the two spin-polarized parabolas cannot be properly resolved. Second, the photoemission intensity of the $sp_z$ state for small $x$ is suppressed for $k>0$  at $h\nu=21.2$\,eV (see Fig. \ref{fig:ARPES}a) due to photoemission matrix element effects, so that only part of the bands are available for the fitting procedure described above which limits the accuracy of the fits. However, the main trend --- a continuous increase of the spin splitting with Bi content $x$ --- is clearly reproduced by the SARPES data.

A more elaborate analysis of the spin-resolved data allows us to determine all three components of the photoelectron spin polarization as explained in Ref. \cite{Meier}. For small $x$, when the spin splitting is comparable to the momentum broadening, a spin state interference in the photoemission process is observed in the region where states with orthogonal spinors overlap. This effect creates an artificial out-of-plane spin-polarization that does not reflect the spin-polarization of the initial state \cite{Meier2}. However, this spin state interference only takes place for small $x$ and can be well separated from the spin-polarization resulting from the Rashba effect. We find that, in contrast to the size of the spin splitting, the spin-direction of the initial state does not depend significantly on $x$ in agreement with previous results \cite{Meier1}.

\begin{figure}
  \includegraphics[width = 1\columnwidth]{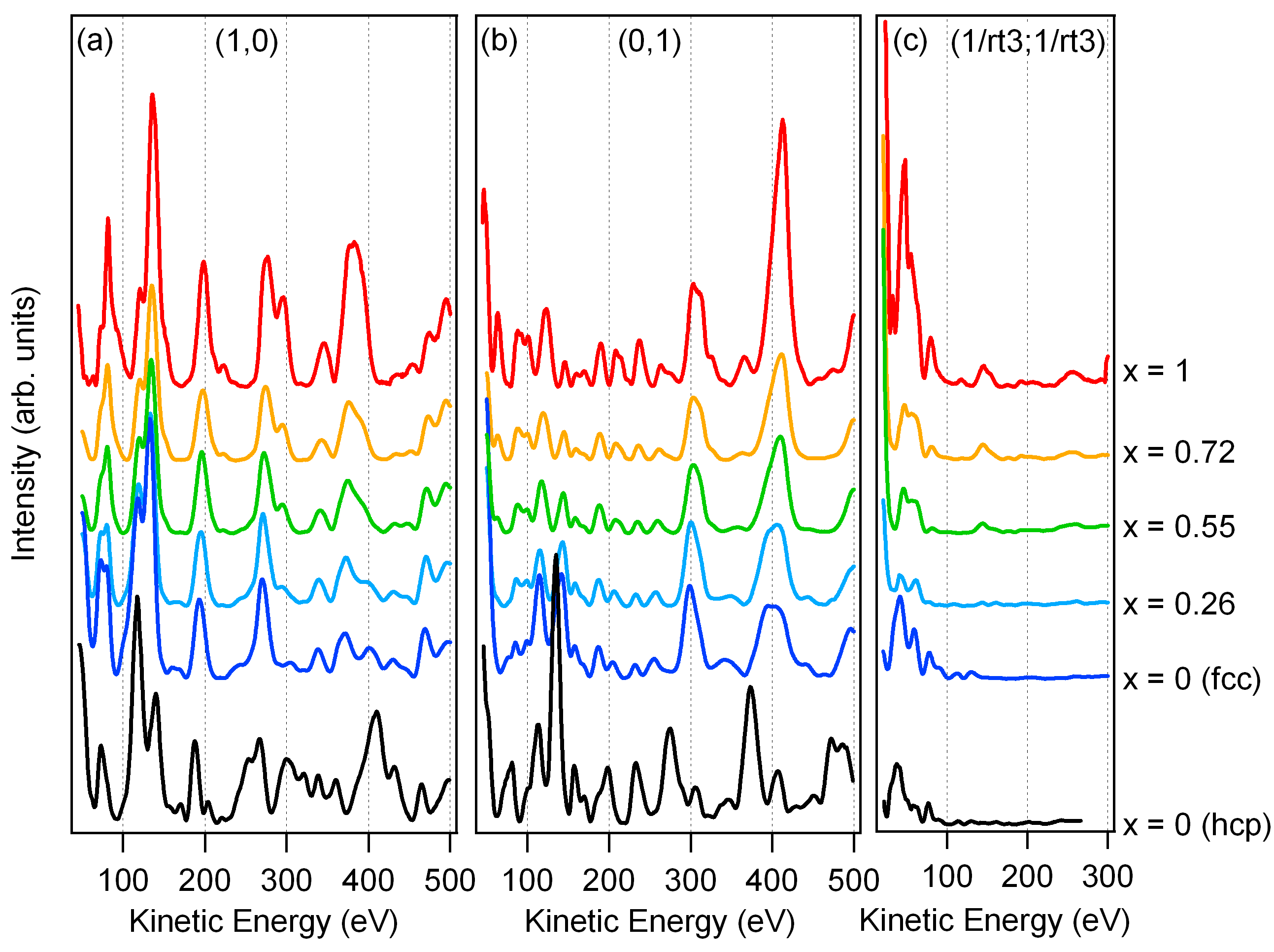}
  \caption{(Color online) Low-energy electron diffraction from mixed Bi$_{x}$Sb$_{1-x}$/Ag(111) surface alloys. The $I(V)$ spectra represent the integrated intensities of the (1,0) (a), (0,1) (b) and ($1/\sqrt{3}$, $1/\sqrt{3}$) (c) spots versus electron energy. The spectra evolve continuously between $x = 0$ and $x = 1$. To exclude a possible hcp toplayer stacking, spectra for hcp stacked Sb/Ag(111) are displayed for comparison (bottom).}
  \label{fig:LEED}
\end{figure}

The ($\sqrt3\times\sqrt3$)R30$^{\circ}$ phase of Sb/Ag(111) can be formed with either fcc or hcp toplayer stacking \cite{Gierz,Woodruff,Quinn}. The faulted hcp toplayer stacking of the Sb/Ag(111) surface alloy is accompanied by the presence of subsurface stacking faults in the Ag substrate caused by Sb diffusion into the bulk \cite{Woodruff}. In experiment, the toplayer stacking can be controlled by tuning the energy of the deposited Sb ions during Sb deposition with an electron beam evaporator. The Sb/Ag(111) surface alloys with fcc and hcp toplayer stacking, respectively, can easily be identified with the help of $I(V)$-LEED measurements, where the intensity $I$ of a particular diffraction spot is measured as a function of the kinetic energy of the incident electrons which is controlled by the acceleration voltage $V$ \cite{Gierz}.

In order to determine the toplayer stacking for the mixed Bi$_{x}$Sb$_{1-x}$/Ag(111) surface alloy, we investigated the surface structure with $I(V)$-LEED (Fig. \ref{fig:LEED}). The $I(V)$-LEED spectra were averaged over equivalent spots and smoothed (further details are given in \cite{Gierz}). They evolve continuously between the pure Bi/Ag(111) surface alloy (red in Fig. \ref{fig:LEED}) and the pure Sb/Ag(111) surface alloy with fcc top layer stacking (blue). For comparison, the $I(V)$ spectra for the Sb/Ag(111) surface alloy with hcp top layer stacking are shown in black. As these spectra clearly differ from those for the mixed Bi$_{x}$Sb$_{1-x}$/Ag(111) surface alloy, we conclude that it forms with fcc top layer stacking even if Sb ions are deposited at sufficiently high energies to form an hcp stacked surface alloy.

\subsection{Ternary surface alloy Bi$_{x}$Pb$_{y}$Sb$_{1-x-y}$/Ag(111)}

As was shown before, both Bi$_x$Pb$_{1-x}$/Ag(111) and Bi$_{x}$Sb$_{1-x}$/Ag(111) can be formed and exhibit a well-defined band structure. While in the Bi$_x$Pb$_{1-x}$/Ag(111) surface alloy both the spin splitting and the Fermi level change as a function of $x$, it is possible to change the spin splitting while leaving the Fermi level largely unaffected in the corresponding Bi$_{x}$Sb$_{1-x}$/Ag(111) surface alloy. As transport properties are determined by the spin texture of the Fermi surface it is important to tune both the position of the Fermi level as well as the size of the spin splitting independently. This can be achieved in a ternary Bi$_{x}$Pb$_{y}$Sb$_{1-x-y}$/Ag(111) surface alloy as follows. In order to increase $k_R$ one has to add heavy elements with large outward relaxation (i.e. Bi, Pb). For a modification of $E_F$ Bi$_{x}$Sb$_{1-x}$ should be mixed with Pb.

\begin{figure}
  \includegraphics[width = 1\columnwidth]{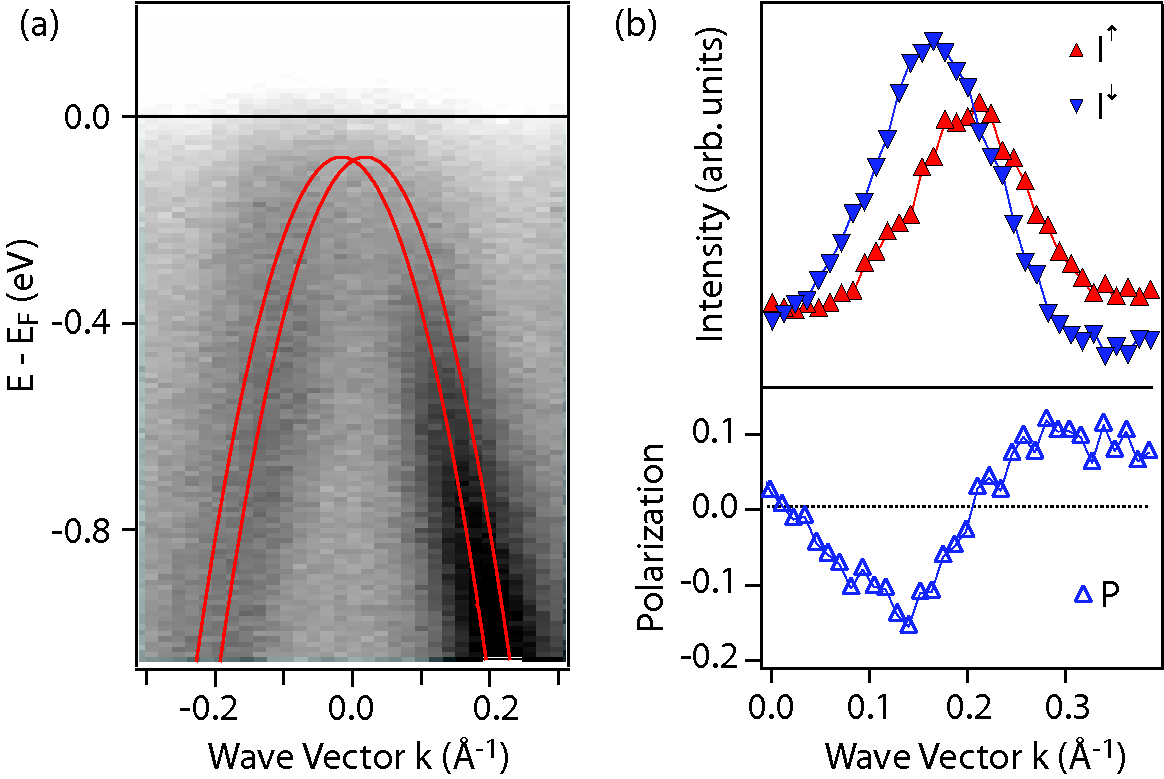}
  \caption{(Color online) Spin-resolved photoemission from the ternary Bi$_{0.3}$Pb$_{0.35}$Sb$_{0.35}$/Ag(111) surface alloy. (a) Surface-state dispersion around $\bar{\Gamma}$. Red lines are guides to the eye. (b) Spin-resolved intensities $I^{\uparrow}$ and $I^{\downarrow}$ and spin polarization $P$ measured at an initial state energy of -0.75\,eV. The spin splitting of the $sp_z$ surface state is clearly resolved.}
  \label{fig:SARPES2}
\end{figure}

Figure \ref{fig:SARPES2}a shows the surface state band structure for a mixed ternary surface alloy with $(x, y, z=1-x-y) = (0.3, 0.35, 0.35)$. The line width is significantly increased as compared to those for the mixed binary surface alloys. The ARPES measurements in Fig. \ref{fig:SARPES2}a were done using the COPHEE spectrometer with a reduced angular resolution as compared to the setup used for the ARPES data in Fig. \ref{fig:ARPES}. An additional increase in line width is caused by the intermixing of three elements (instead of two) and by a --- probably --- not optimized sample preparation. Despite the large line width, the spin splitting ($\Delta k/2=0.019$\,\AA$^{-1}$) can still be clearly resolved with SARPES, as shown in Fig. \ref{fig:SARPES2}b.

\section{Conclusions}

We have shown that it is possible to form the mixed binary Bi$_x$Sb$_{1-x}$/Ag(111) surface alloy despite the fcc/hcp top layer stacking of Sb/Ag(111). Furthermore, $I(V)$-LEED experiments revealed that the mixed Bi$_x$Sb$_{1-x}$/Ag(111) surface alloy always forms with fcc top layer stacking. ARPES and SARPES measurements show a continuous evolution of the band structure with Bi content $x$. The results on the mixed ternary surface alloy Bi$_{0.3}$Pb$_{0.35}$Sb$_{0.35}$/Ag(111) show that ternary alloys can be formed and exhibit a reasonably well defined band structure. Our findings indicate the possibility to form ternary surface alloys with arbitrary compositions that allow us to tune the spin splitting as well as the Fermi energy independently and continuously over a broad range of values.

Following this idea, the situation where the Fermi level lies in between the band maximum and the crossing point of the two parabolas is particularly interesting: in this regime the spins on the two circular Fermi surfaces rotate in the same direction and the density of states shows quasi one-dimensional behavior \cite{Ast2}. In this case, the Rashba energy becomes the dominating energy scale and an increase of the transition temperature into a superconducting state is expected \cite{Cappelluti}. Furthermore, mixed surface alloys with a spatial variation of their chemical composition $x$ create a gradient of the effective Rashba field that is a prerequisite for building a Stern-Gerlach spin filter \cite{Ohe}.

\end{document}